\documentclass[preprintnumbers, floatfix, showkeys, preprintnumbers, letterpaper, twocolumn, superscriptaddress,nofootinbib]{revtex4-2}
\pdfoutput=1 
\usepackage{graphicx}
\usepackage{microtype}
\usepackage{amsmath}
\usepackage{amssymb}
\usepackage{subfigure}
\usepackage{url}
\usepackage{hyperref}
\usepackage{mathtools}
\usepackage{orcidlink}

\usepackage{xcolor}
\usepackage{color}
\usepackage{mathrsfs}
\usepackage{calrsfs}
\usepackage{amsfonts}
\usepackage{tabularx}
\usepackage{latexsym}
\usepackage{ragged2e}
\usepackage{epsfig}
\usepackage{textcomp}
\usepackage{float}

\usepackage{caption}
\DeclareCaptionJustification{justified}{\leftskip=0pt \rightskip=0pt \parfillskip=0pt plus 1fil}
\definecolor{vividviolet}{rgb}{0.62, 0.0, 1.0}
\definecolor{amaranth}{rgb}{0.9, 0.17, 0.31}
\definecolor{palatinateblue}{rgb}{0.15, 0.23, 0.89}
\definecolor{brightpink}{rgb}{1.0, 0.0, 0.5}
\definecolor{cornflowerblue}{rgb}{0.39, 0.58, 0.93}
\definecolor{deepcarminepink}{rgb}{0.94, 0.19, 0.22}
\definecolor{radicalred}{rgb}{1.0, 0.21, 0.37}

\hypersetup{ linktoc=all,
	colorlinks, linkcolor={palatinateblue},
	citecolor={brightpink}, urlcolor={amaranth}
}

\graphicspath{{Images/}}

\def\sideremark#1{\ifvmode\leavevmode\fi\vadjust{\vbox to0pt{\vss
			\hbox to 0pt{\hskip\hsize\hskip1em
				\vbox{\hsize1.3cm\tiny\raggedright\pretolerance10000
					\noindent #1\hfill}\hss}\vbox to8pt{\vfil}\vss}}}%
\def\beq{\begin{equation}}
\def\eeq{\end{equation}}

\begin{document}
\title{Do Black Holes With Generalized Entropy Violate Bekenstein Bound?}

\author{Hengxin \surname{L\"u}}
\email{mx120220337@stu.yzu.edu.cn}
\affiliation{Center for Gravitation and Cosmology, College of Physical Science and Technology,\\ Yangzhou University,
 Yangzhou, 225002, China}%

\author{Yen Chin \surname{Ong}\orcidlink{0000-0002-3944-1693}}
\email{ycong@yzu.edu.cn}
\affiliation{Center for Gravitation and Cosmology, College of Physical Science and Technology,\\ Yangzhou University,
 Yangzhou, 225002, China}
\affiliation{Shanghai Frontier Science Center for Gravitational Wave Detection, School of Aeronautics and Astronautics,\\ Shanghai Jiao Tong University, Shanghai 200240, China}

\begin{abstract}
In general yes, but also not quite. It is known that if the Bekenstein-Hawking entropy is replaced by some kind of generalized entropy, then the Bekenstein bound may be grossly violated. In this work, we show that this undesired violation can be avoided if we employ the equivalence between generalized entropy and varying-$G$ gravity (GEVAG). In this approach, modifying entropy necessarily also modifies gravity (as one should expect if gravity is indeed inherently tied to thermodynamics), which leads to an effective gravitational ``constant'' $G_\text{eff}$ that is area-dependent, and a thermodynamic energy that is distinct from the ADM mass. We show that a relaxed Bekenstein bound of the form $S \leqslant CRE$ is always satisfied, albeit the coefficient $C$ is no longer $2\pi$.    
\end{abstract}

\maketitle

\section{Introduction: Generalized Entropy and Bekenstein Bound}\label{I}

A system with entropy $S$, a suitably defined size $R$ and an energy $E$ satisfies
the standard Bekenstein bound \cite{Bekenstein}  
\begin{equation}
S \leqslant \frac{2\pi k_B}{\hbar c}RE = 2\pi RE,
\end{equation}
where, as per the common practice in literature, we have set $\hbar = c = k_B =1$. We will keep $G$ explicit since we will discuss modification of the gravitational constant later.
In general relativity, we can easily check that a Schwarzschild black hole saturates this bound. The Bekenstein bound also holds in non-gravitational system (not surprising --- note the absence of $G$ in the inequality), and is related to the positivity of the relative entropy \cite{0804.2182,2409.14408}. The quantum informational perspective of the Bekenstein bound is still an active area of research \cite{2309.07436}. The generality of the bound suggests that it could be fundamental, at least in asymptotically flat spacetimes\footnote{Its status in anti-de Sitter spacetimes are less clear. See \cite{0606186}.}. On the other hand, in the literature, one finds many attempts to study what happens if we were to replace Bekenstein-Hawking entropy with various generalized entropies, such as Barrow entropy \cite{2004.09444},
Tsallis(-Cirto) entropy \cite{Tsallis,Tsallis2}, 
Kaniadakis entropy \cite{0210467,0507311,2109.09181}, 
and Sharma-Mittal entropy \cite{1802.07722}, as well as generalizations thereof \cite{2201.02424}. 

One of the motivations for contemplating these other possibilities is that the Bekenstein-Hawking entropy $S={A}/{4G}=4\pi G^2 M^2$ is not extensive in its energy $M$ (that is, $S(\lambda M)\neq \lambda S(M)$), which means that its underlying statistics is probably not the Gibbs-Boltzmann entropy. As the logic goes, since the Bekenstein-Hawking entropy is not extensive anyway, one might as well consider other non-extensive forms that were first considered in non-gravitational systems, and use it to generalize the Bekenstein-Hawking form. (The Barrow case is somewhat special as it was motivated as a quantum gravity effect.) This certainly does not look like a strong motivation, and so we do not advocate for the applications of generalized entropy for black holes. While we need not believe in any of these extensions, it is important to study the effects of modifying Bekenstein-Hawking entropy on black hole physics. If the Bekenstein-Hawking entropy is in fact correct and therefore unique, then perhaps we can see what will go wrong by modifying it. (Similarly, modified gravity theories also often lead to subtle consistency issues.) 

One thing that could potentially go \emph{very} wrong is the Bekenstein bound. To see this, we consider for example, the Tsallis entropy, usually given in the context of black holes as\footnote{This may not be the case, see, e.g., \cite{2502.02522,2505.03061}.}
\begin{equation}\label{st}
S_T \coloneq \frac{A_0}{4G}\left(\frac{A}{A_0}\right)^\delta = \frac{A_0}{4G}\left(\frac{4\pi R^2}{A_0}\right)^\delta,
\end{equation}
where $A_0$ is a constant of the theory.
The right hand side (RHS) of the Bekenstein bound is
\begin{equation}
2\pi RE = 2\pi R M = \frac{\pi R^2}{G}.
\end{equation}
The bound is only satisfied if
\begin{equation}
S_T^{1-\frac{1}{\delta}} \leqslant \left(\frac{A_0}{4G}\right)^{1-\frac{1}{\delta}}=\text{const.}
\end{equation}
Evidently, since for fixed $\delta$, the Tsallis entropy $S_T$ is an increasing function of $A$, we see that the bound fails for \emph{large} black holes. This was first pointed out in \cite{2207.13652} (the example we use here is similar to their computation for the Barrow case).
In general then, as argued in \cite{2207.13652,2411.00694}, when a generalized entropy is used in place of Bekenstein-Hawking one, we should not expect the Bekenstein bound to hold. 
Thus, if we take the position that the Bekenstein bound is fundamental, this would either imply that generalized entropies are not viable (and thus this can be used to constrain parameters of the theory; see e.g.,\cite{2405.14799}), or that we \emph{cannot} naively directly apply any generalized entropy with the usual assumption that $R=2GM$ and $E=M$. In fact, changing the entropy alone while keeping everything else unchanged has been long realized to be problematic and may be inconsistent (see, e.g., \cite{2109.05315,2207.07905}; also see \cite{2307.01768} for other criticisms in the context of cosmology), though this is still widely done in the literature.

In \cite{2407.00484}, we pointed out that if we were to take the claim that gravity is deeply related to thermodynamics seriously, then it is only natural that gravity will also need to be modified when we generalize the Bekenstein-Hawking entropy. We proposed that one possible way to do this is via generalizing Jacobson's method in deriving the Einstein field equations \cite{J}. Specifically, we showed that if we generalized the Bekenstein Hawking entropy by changing the area $A$ to some (dimensionful) function $f(A)$, namely
\begin{equation}
\frac{A}{4G} \longmapsto\frac{f(A)}{4G},
\end{equation}
then the resulting theory has a field equation that has the same form as general relativity, except that its gravitational constant $G$ has to be replaced by 
\begin{equation}\label{GEFF}
G\longmapsto G_\text{eff} \coloneq \frac{G}{f'(A)},
\end{equation}
where $A$ is the area of the black hole horizon, and prime denotes derivative with respect to $A$. 
Since $G_\text{eff}$ is in general a function unless the horizon is fixed, it is a novel class of varying-$G$ theory.
We shall refer to this approach as the GEVAG approach, short for ``\textbf{G}eneralized \textbf{E}ntropy and \textbf{VA}rying-$\mathbf{G}$ gravity''. In this work, because we only consider the Schwarzschild (static) solution, $G_\text{eff}$ is fixed and we need not worry about its variation. Since the field equation of GEVAG has the same form as general relativity,
the Schwarzschild solution has event horizon located at $r=2 G_\text{eff} M$. For more details, we refer the readers to \cite{2407.00484}.

This area dependence looks strange and suspicious, and we spent a great deal of effort making sense of it in \cite{2407.00484}. Still, this claim has an unexpected supporting evidence. If we consider the usual logarithmic quantum gravity correction to the Bekenstein-Hawking entropy
\begin{equation}
S = \frac{A}{4G} + \tilde{c} \ln\left(\frac{A}{G}\right),
\end{equation}
for some constant $\tilde{c}$, then applying the result above, we have
\begin{equation}\label{Geff}
G_\text{eff} = \frac{G}{1+\dfrac{\tilde{c}G}{A}}. 
\end{equation}
This is exactly the form one expects in the asymptotically safe gravity scenario:
\begin{equation}\label{ASG}
G(k)= \frac{G({k_0})}{1+ \mathfrak{c}k^2},
\end{equation}
where $k_0$ is a reference energy scale and $\mathfrak{c}$ another constant. In fact,  the authors in \cite{2204.09892} (see also \cite{2308.16356}) had argued that $k$ is horizon area dependent: $k = \text{const.}/\sqrt{A}$. This implies Eq.(\ref{ASG}) is exactly the same as Eq.(\ref{Geff}). In other words, our GEVAG approach unexpectedly implies a connection between the standard logarithmic correction and the ASG approach. It is also interesting to note that the area dependence disappears when the entropy is linear in $A$ only. Thus the Bekenstein-Hawking entropy is in a special class (unfortunately this does not fix the constant prefactor to be $1/4$).

Therefore, to check whether the Bekenstein bound is really consistent with generalized entropy, we need to check it in the GEVAG approach. In fact, based on the results already obtained in\footnote{There are some problems and unclarified subtleties in \cite{2207.09271}, so the results in \cite{2407.00484} superseded it.} \cite{2207.09271}, we have claimed in \cite{2407.00484} that the Bekenstein bound is satisfied, possibly up to some constant prefactor. However, since Bekenstein bound wasn't the main focus therein, we did not elaborate on the details. In this work, we shall explicitly demonstrate this to be indeed the case with two explicit examples: Tsallis entropy and R\'enyi entropy. We then prove the general result for any generalized entropy $S=f(A)/4 G$.

\section{Bekenstein Bound For Tsallis-Schwarzschild Black Hole}

In the GEVAG approach, the gravitational constant for the Tsallis entropy case becomes, via Eq.(\ref{GEFF}),
\begin{equation}\label{geff0}
G_\text{eff}=\frac{G}{\delta}\left(\frac{A}{A_0}\right)^{1-\delta}.
\end{equation}
Furthermore, via the first law $dE=TdS$, where $T=1/{8\pi G_\text{eff}M}$,
one can show that the thermodynamic energy is not the same as the ADM mass $M$, but rather \cite{2407.00484}
\begin{equation}\label{ETsallis}
E=\frac{M}{2\delta -1}.
\end{equation}
As emphasized in \cite{2407.00484}, this relation is much simpler than what one would obtain from assuming $E\neq M$, but without the GEVAG consideration \cite{2106.00378,2207.09271}. In addition, positivity of energies in Eq.(\ref{ETsallis}) imposes the bound $\delta > 1/2$.

We now note that 
\begin{equation}\label{st}
S_T =  \frac{A_0}{4G}\left(\frac{4\pi R^2}{A_0}\right)^\delta = \frac{A_0}{4G}\left[\frac{4\pi (2G_\text{eff}M)^2}{A_0}\right]^\delta.
\end{equation}
On the other hand, Eq.(\ref{geff0}) yields a relationship between $G_\text{eff}$ and $G$:
\begin{equation}
G=\frac{M^{2(\delta-1)}(16\pi)^{\delta-1}A_0^{1-\delta}\delta}{G_\text{eff}^{1-2\delta}}.
\end{equation}
We can substitute this into Eq.(\ref{st}) and obtain
\begin{equation}
S=\frac{4\pi}{\delta} G_\text{eff} M^2.
\end{equation}
On the other hand, the RHS of the Bekenstein bound is 
\begin{equation}
2\pi RE = 2\pi (2G_\text{eff}M)\frac{M}{2\delta-1}=\frac{4\pi}{2\delta-1}G_\text{eff}M^2.
\end{equation}
Therefore, $S/RE \leqslant 2\pi$ holds if and only if
\begin{equation}
\frac{1}{\delta} \leqslant \frac{1}{2\delta-1} \Longleftrightarrow \delta \geqslant 2\delta-1,
\end{equation}
and thus for all $\delta \leqslant 1$. Otherwise, we can consider the ``relaxed form'' of Bekenstein bound by replacing the constant $2\pi$ with some constant $C$. Then the relaxed Bekenstein bound
\begin{equation}
S \leqslant CRE
\end{equation}
is satisfied for $S_T$ if we take the ``Bekenstein constant'' $C$ to be
\begin{equation}\label{Ctsa}
C= 2\pi\left(2-\frac{1}{\delta}\right) \leqslant 4\pi.
\end{equation}
Some clarifications are useful at this point. Let us first return to the original Bekenstein bound. The ratio $S/RE$ is strictly less than $2\pi$ for non-black hole systems, and is saturated by a black hole. In principle, $S/RE\leqslant C$ for any $C\geqslant 2\pi$ would also be mathematically correct, but physically $C> 2\pi$ would correspond to a hypothetical ``hyper-entropic'' object (i.e. an object whose entropy exceeds a black hole of the same energy). Such an object is likely to collapse into a black hole \cite{0908.1265,0706.3239v2, 1304.3803,2201.11132}. In fact, the Bekenstein bound means that such objects should be excluded --- they would collapse into a black to restore the Bekenstein bound. For the same reason\footnote{Admittedly there is a caveat: whether hyper-entropic objects are indeed unstable for the case of generalized entropy requires more detailed investigations.}, we \emph{define} the ``Bekenstein constant'' $C$ as the value that is saturated by a black hole in the case with generalized entropy.

\begin{figure}[!h]
\centering
\includegraphics[width=0.50\textwidth]{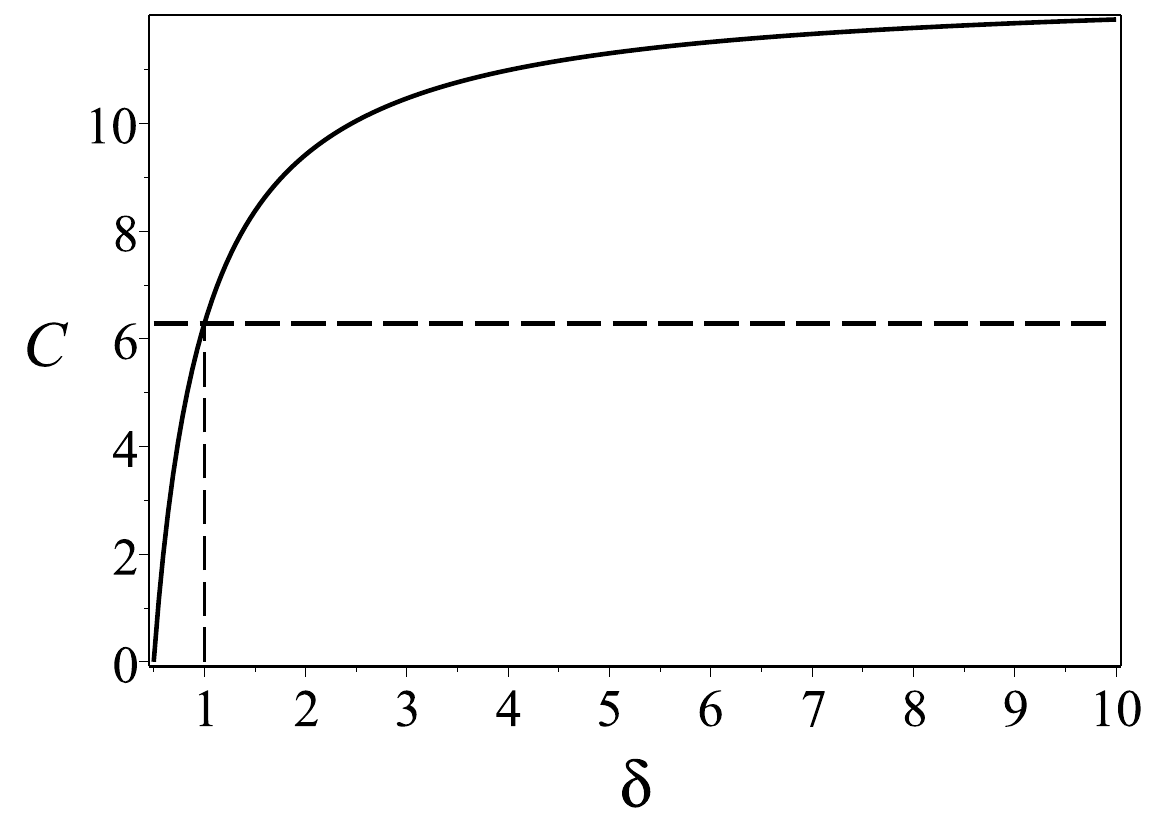}
\caption{The Bekenstein constant $C$ for the Tsallis-Schwarzschild black hole is bounded between 0 and $4\pi$, with $0< C \leqslant 2\pi$ for $1/2 <\delta \leqslant 1$. The dashed vertical line and horizontal line are $\delta=1$ and $C=2\pi$, respectively. \label{tsallis}}
\end{figure}

Eq.(\ref{Ctsa}) is an increasing function that reduces to $2\pi$ when $\delta \to 1$, consistent with the standard Bekenstein bound for Bekenstein-Hawking entropy. In fact, we can see from Fig.(\ref{tsallis}) that the original Bekenstein bound is satisfied for $\delta =1$, and that $C$ is strictly less than $2\pi$, indicating an even tighter bound than the original version, provided that $1/2 <\delta \leqslant 1$. In the limit of large $\delta$, we have $C \to 4\pi$, twice the standard value. 

This suggests that the relaxed Bekenstein bound can also be classified as:
\begin{itemize}
\item[(1)] \underline{Strong Bekenstein Bound:} $0<C < 2\pi$,
\item[(2)] \underline{(Original) Bekenstein Bound:} $C=2\pi$,
\item[(3)] \underline{Weak Bekenstein Bound:} $2\pi < C$, with $C/2\pi = O(1)$.
\end{itemize}

Note that the strong bound is not necessarily good if $C$ is too small. This is because one interpretation of the Bekenstein bound is that if a system has too much entropy it will collapse into a black hole, thus restoring the bound. If $C$ is too small, this suggests black hole production is much easier, which will likely lead to conflicts with the observed number density of black hole of various masses.  

\section{Bekenstein Bound For R\'enyi-Schwarzschild Black Hole}

We now consider another form of entropy that has been widely considered in the literature: the R\'enyi entropy \cite{renyi}. For black holes, this take the form \cite{1309.4261,1511.06963}
\begin{equation}\label{renyi}
S_R \coloneq \frac{1}{\alpha}\ln\left[1+\alpha\left(\frac{A}{4G}\right)\right], ~\alpha>0,
\end{equation}
for which the limit $\alpha \to 0$ recovers Bekenstein-Hawking entropy.

In the GEVAG approach, the gravitational constant for the R\'enyi case becomes, via Eq.(\ref{GEFF}),
\begin{equation}\label{GEFF2}
G_\text{eff} = \frac{G}{f'(A)}=G\left(1+\frac{1}{4}\frac{\alpha A}{G}\right).
\end{equation}
This can be solved in terms of $G$ only by substituting in the Schwarzschild radius $r=2G_\text{eff}M$. The result is
\begin{equation}\label{geff}
G_\text{eff} = \frac{1-\sqrt{1-16\pi \alpha GM^2}}{8\pi \alpha M^2}.
\end{equation}
Thus the GEVAG approach readily implies a non-obvious upper bound for the R\'enyi parameter, which is crucial to ensure that $G_\text{eff} \in \Bbb{R}$: 
\begin{equation}
\alpha \leqslant \frac{1}{16\pi G M^2}.
\end{equation}
This bound was also found in \cite{2504.16705}, although with a different approach and assumptions.

Note that Eq.(\ref{GEFF2}) implies 
\begin{equation}\label{S_R}
S_R \coloneq \frac{1}{\alpha}\ln\left[1+\alpha\left(\frac{A}{4G}\right)\right] = \frac{1}{\alpha}\ln\left(\frac{G_\text{eff}}{G}\right).
\end{equation}
Let us now derive the thermodynamic energy. The first law is $dE=T dS_R$. The Hawking temperature of the black hole is $T=1/8\pi G_\text{eff}M$ \cite{2407.00484}. Substituting in the expression for $S_R$ in Eq.(\ref{S_R}), we can use the chain rule to get
\begin{equation}
dE =T\frac{dS_R}{dM} dM.
\end{equation}
Since 
\begin{flalign}
\frac{dS_R}{dM} &= \frac{1}{\alpha}\frac{d}{dM}\left[\ln\left[1+\alpha\left(\frac{4\pi(2G_\text{eff} M)^2}{4G}\right)\right]\right] \\ \notag
&= \frac{1}{\alpha}\left(\frac{1-\sqrt{1-16\pi \alpha G M^2}}{M\sqrt{1-16\pi \alpha G M^2}}\right),
\end{flalign}
upon simplifying the expression we finally obtain
\begin{equation}
dE = \frac{dM}{\sqrt{1-16\pi \alpha GM^2}}. 
\end{equation}
Integrating this finally gives the thermodynamic energy:
\begin{equation}
E = \frac{1}{4\sqrt{\pi\alpha G}} \arctan\left(\frac{4\sqrt{\pi \alpha G}M}{\sqrt{1-16\pi \alpha G M^2}}\right),
\end{equation}
which is considerably more complicated than the Tsallis case. Let us define the dimensionless quantity 
\begin{equation}
a\coloneq \alpha GM^2,
\end{equation}
so that the energy-mass relation can be written as
\begin{equation}
E=M\left[\frac{1}{\sqrt{16\pi a}} \arctan\left(\frac{\sqrt{16\pi a}}{\sqrt{1-16\pi a}}\right)\right].
\end{equation}
We note that
\begin{equation}
\lim_{a\to {1/16\pi}}\left[\arctan\left(\frac{\sqrt{16\pi a}}{\sqrt{1-16\pi a}}\right)\right]=\frac{\pi}{2},
\end{equation}
and 
\begin{equation}
\lim_{a\to {1/16\pi}}\frac{M}{\sqrt{16\pi a}}=M.
\end{equation}
Therefore
\begin{equation}
\lim_{a\to {1/16\pi}} E =\frac{\pi M}{2}.
\end{equation}

\begin{figure}[!h]
\centering
\includegraphics[width=0.50\textwidth]{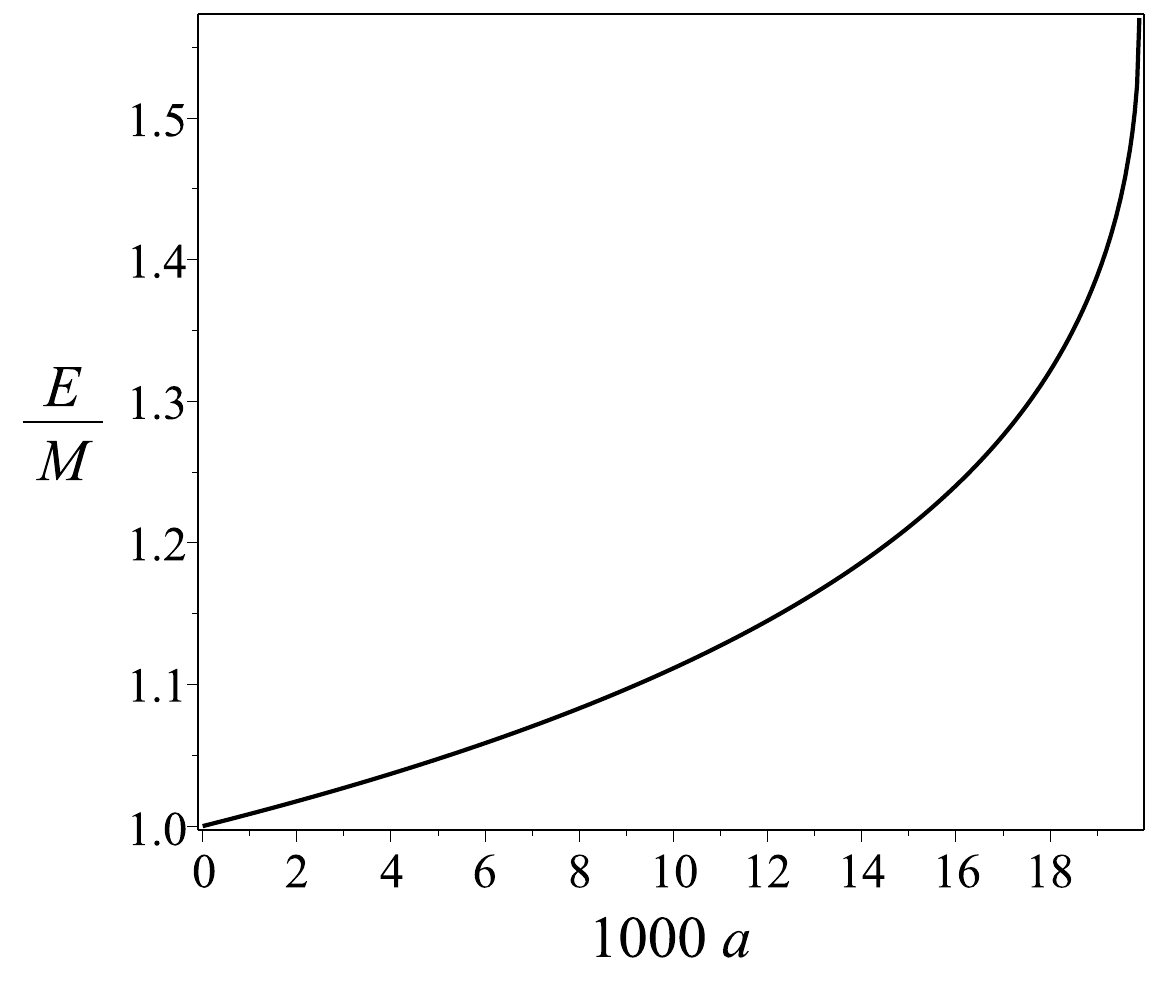}
\caption{The ratio $E/M$ for the R\'enyi-Schwarzschild  black hole. The horizontal axis has been re-scaled for clarity. The domain of $a$ is $0 < a < 1/16\pi \approx 0.0199$. We have $E/M=1$ when $a=0$, the GR case. On the other hand, $E/M$ tends to $\pi/2$ when $a\to 1/16\pi$. \label{renyimass}}
\end{figure}

Thus the ratio $E/M \to \pi/2$ in the limit $a \to 1/16$.
The behavior of $E/M$ is shown in Fig.(\ref{renyimass}).
In the small $a$ expansion, the thermodynamic energy $E$ reduces to
\begin{equation}
E = M + \frac{8}{3}a{\pi}M + \cdots,
\end{equation}
which smoothly reduces to $E=M$ in the limit $a \to 0$. 

Again, we want to find the $C$ in the relaxed Bekenstein bound $S \leqslant C RE$. Since $C$ is defined to be the saturated value of the generalized Bekenstein bound, it can be obtained as $C=S/RE$.
A direct computation yields 
\begin{equation}
C = \dfrac{\dfrac{1}{\alpha}\ln\left[1+\alpha\left(\dfrac{A}{4G}\right)\right]}{2G_\text{eff}M \left[\dfrac{1}{4\sqrt{\pi\alpha G}} \arctan\left(\dfrac{4\sqrt{\pi \alpha G}M}{\sqrt{1-16\pi \alpha G M^2}}\right)\right]}.
\end{equation}
Substituting in $A=4\pi(2G_\text{eff}M)^2$ and the expression for $G_\text{eff}$ in Eq.(\ref{geff}), we finally obtain in terms of $a$,
\begin{equation}
C =\frac{16 \sqrt{a\pi^3}\ln\left(\dfrac{1-\sqrt{1-16\pi a}}{8\pi a}\right)}{\left(1-\sqrt{1-16\pi a}\right)\arctan\left(\dfrac{4\sqrt{\pi a}}{\sqrt{1-16\pi a}}\right)}.
\end{equation}

As shown in Fig.(\ref{renyi}), unlike the Tsallis case, this is a decreasing function, so it always satisfies $C \leqslant 2\pi$. 
In this case the strong Bekenstein bound is strictly satisfied.
In the limit $a \to 0$, we recover the standard value $2\pi$ for the Bekenstein constant.
It should be emphasized that it is crucial to make the distinction between the strong Bekenstein bound and the standard Bekenstein bound. In the latter, black holes saturate the bound, while other systems satisfy strict inequality. Whereas for the former, \emph{black holes} saturate a bound whose value $C$ is less than $2\pi$. Due to this physical difference, it would be misleading to say that such generalized entropy black holes satisfy the original bound.

\begin{figure}[!h]
\centering
\includegraphics[width=0.50\textwidth]{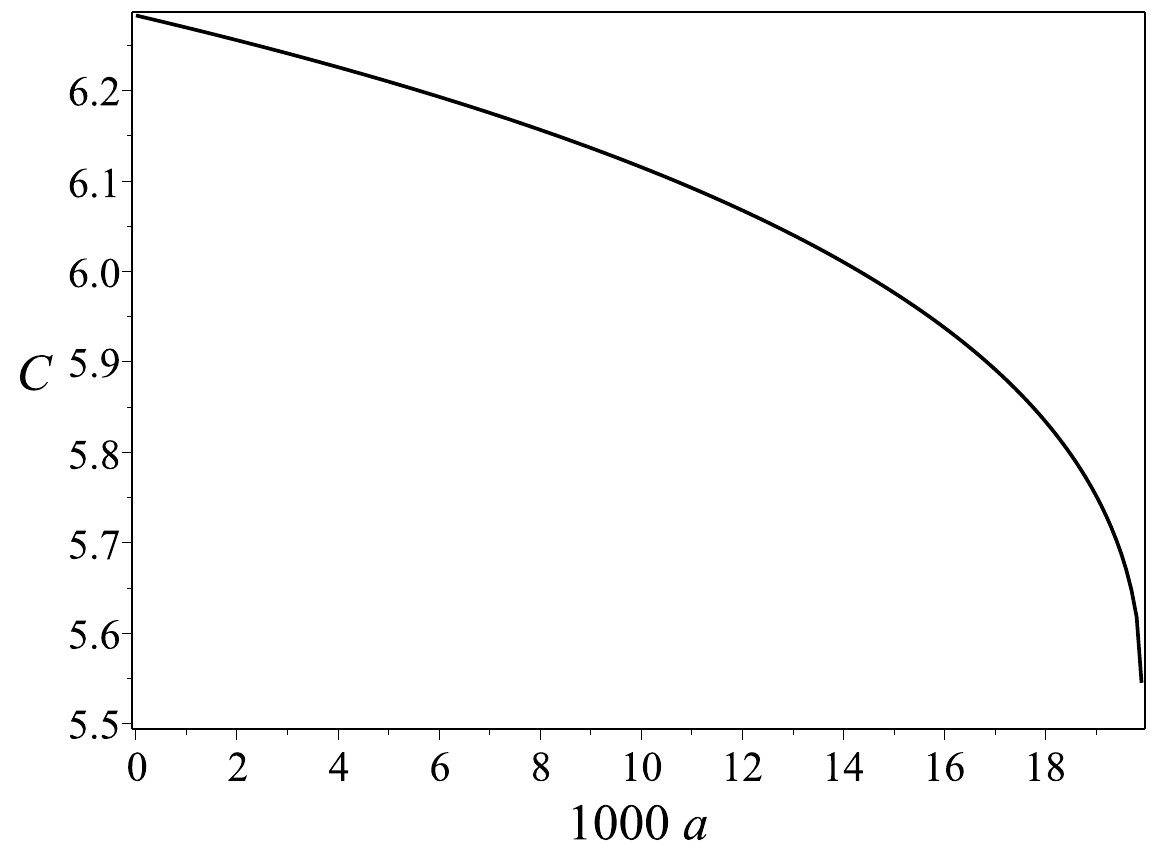}
\caption{The Bekenstein constant $C$ for R\'enyi-Schwarzschild black hole, as a function of the dimensionless R\'enyi parameter $a \coloneq \alpha GM^2$. The horizontal axis has been re-scaled for clarity. The domain of $a$ is $0 < a < 1/16\pi \approx 0.0199$. We have $C \to 2\pi$ in the limit $a\to 0$, and $C \to 8 \ln 2 \approx 5.5452$ when $a\to 1/16\pi$. \label{renyi}}
\end{figure}

\section{Conclusion: General Result}

Given the vast literature of black holes with generalized entropy, it is crucial to clarify whether such black holes violate the Bekenstein bound or not. If one applies the generalized entropy directly, Bekenstein bound is indeed not expected to hold in general. However, instead of ruling out such generalized entropies, this could just be a hint that changing the entropy while keeping everything else unchanged is inconsistent. In this work, utilizing the GEVAG approach, we have explicitly demonstrated that both the Tsallis(-Cirto)-Schwarzschild black hole and the R\'enyi-Schwarzschild black hole satisfy the Bekenstein bound up to a constant term, as we claimed in \cite{2407.00484}.

Let us now present the general result for any generalized entropy $S=f(A)/4G$. For this to satisfy $S \leqslant CRE$, with $R=2G_\text{eff} M$ {for the Schwarzschild black hole under GEVAG scheme \cite{2407.00484}}, it suffices to let the Bekenstein constant be
\begin{equation}\label{C}
C = \frac{f(A)}{8GG_\text{eff}ME}.
\end{equation}

This combination is clearly dimensionless and reduces to $2\pi$ in the GR limit. Furthermore, and crucially, we see that $C$ is bounded because $f(A)$ should be bounded by assumption (otherwise the generalized entropy $S=f(A)/4G$ diverges, which does not make physical sense for a finite system). Likewise, $G_\text{eff}=G/f'(A)$ is nonzero as long as the entropy function $f(A)$ is differentiable and $f'(A)$ is finite. This is sensible for the same reason: if $f'(A)$ diverges, then $f$ is becoming unbounded, which is not physical. The energy $E$ should also be finite for any sensible thermodynamical system.
Furthermore, since $E$ should recover $M$ in the GR limit, at least for small deviation away from the Bekenstein-Hawking entropy, $E$ is guaranteed to be finite for any generalized entropy. (If for any unlikely reason, $E$ is divergent but $S$ finite, then the (weak) Bekenstein bound is trivially satisfied.) 

In general, whether $C$ defined by Eq.(\ref{C}) is necessarily less than $2\pi$ has to be checked on a case by case basis.
In our opinion, even if the prefactor is not $2\pi$ but any other finite number larger than $2\pi$, this should not be considered as a true violation of the Bekenstein bound (c.f. the gross violation in Sec.(\ref{I})), since its essential physics that the entropy should be bounded above by the product $RE$ remains intact. Of course, if $C$ is too large, one may start to question whether we want to accept such a theory, but if $C>2\pi$ but $C/2\pi$ is only $O(1)$ or even $O(10)$, such a ``weak Bekenstein bound'' is arguably still reasonable. 

We therefore conclude that the Bekenstein bound, at least a ``relaxed'' form, remains valid if we consistently apply generalized entropy, namely by taking into account the effects of generalized entropy on the theory of gravity itself, via the GEVAG scheme. This sensible results also, in turn, lend credence to the GEVAG approach. Of course, if one insists on requiring that the deviation of $C$ from $2\pi$ is not too large, then one can still use it to constrain the parameter of the theory. 

\end{document}